\documentstyle[prl,epsfig,aps,twocolumn]{revtex}

\begin{document}

\title{Characterization of neutrino signals with radiopulses in dense media 
through the LPM effect}

\author{J. Alvarez-Mu\~niz, R.A. V\'azquez, and E. Zas}

\address{Departamento de F\'\i sica de Part\'\i culas, Universidade de
Santiago E-15706 Santiago de Compostela, Spain}

\maketitle
We discuss the possibilities of detecting radio pulses from 
high energy showers in ice, such as those produced 
by PeV and EeV neutrino interactions. It is shown that the 
rich radiation pattern structure in the 100 MHz to few GHz 
allows the separation of 
electromagnetic showers induced by photons or electrons above 
100~PeV from those induced by hadrons. This opens up the 
possibility of measuring the energy fraction 
transmitted to the electron in a charged current electron 
neutrino interaction with adequate sampling of the 
angular distribution of the signal. 
The radio technique has the potential to complement 
conventional high energy neutrino detectors with 
flavor information.

{\bf PACS number(s):} 96.40.Pq; 96.40.Tv; 95.85.Bh; 13.15.+g 


High energy neutrino detection is one of the 
experimental challenges for the next decade with 
efforts under way to construct large Cherenkov 
detectors arrays under water or ice 
\cite{physreps}. The size of these detectors must 
be in the scale of 1~km$^3$ water equivalent to test  
the neutrino flux predictions above the TeV that arise 
in a number of models attempting to explain the origin 
of highest energy observed cosmic rays and gamma rays \cite{icecube}. 
EeV fluxes are difficult to avoid both in the production 
of the highest energy cosmic rays and in the propagation through the cosmic 
microwave background. Moreover there are allowed regions in 
parameter space for neutrino oscillations \cite{oscillations} which 
may be best probed with high energy neutrinos from cosmological or galactic 
distances \cite{longoscil}. It is thus desirable to explore possibilities for 
alternative neutrino detection such as 
horizontal air showers\cite{Cronin} or radio pulses from high energy 
showers\cite{askaryan,markov}. These techniques may be 
advantageous at sufficiently high energies\cite{price} and can 
provide in any case complementary information relevant for flavor 
identification.  

The detection of coherent radio waves from high energy showers 
has been known since the 60's as an interesting alternative for detecting 
ultra high energy showers \cite{askaryan}. 
These showers develop large excess negative charge because 
the vast majority of the shower particles are in the 
low energy regime dominated by electromagnetic interactions 
with the electrons in the target (Compton, Bhabha, M\"oller 
scattering and electron positron annihilation). The excess 
charge becomes about $20 \%$ of the total number of 
electrons and positrons (shower size), which is proportional 
to shower energy. This excess charge radiates coherently. As 
long as the wavelength is larger than the shower 
dimensions, the electric field amplitude $E=|\vec E|$ 
scales with shower energy. The technique has been proposed 
for detecting neutrino interactions in ice or sand 
\cite{markov}. It has potential advantages such as the 
relatively low cost of the detectors (antennae), 
the large attenuation length for radio waves 
and most importantly the fact that 
information on the excess charge distribution can, 
in principle, be reconstructed from the radiation 
pattern because the radiation is coherent. 
 

When a particle of charge $z$ travels through a medium of refractive 
index $n$ with velocity $\vec {v}=\vec {\beta} c > c/n$ 
Cherenkov light is emitted at the Cherenkov angle $\theta_C$, verifying 
$\cos \theta_C=(\beta n)^{-1}$, with a power 
spectrum given by the well known Frank-Tamm result \cite{fran} : 
\begin{equation}
{d^2 W \over  d\nu d l} = 
\left[{4\pi^2\hbar\over c}\,\alpha\right]z^2\nu
\left[1-{1\over\beta^2 n^2}\right]\;, 
\end{equation}
with $d l= c \beta d t$, the particle track, and $\alpha$ the fine structure 
constant. This is the standard approximation used for most 
Cherenkov applications for wavelengths orders of magnitude
smaller than the tracks. 

The frequency band over which Cherenkov radiation is 
emitted can extend well beyond the familiar optical band if 
the medium is transparent. 
As the radiation wavelength becomes comparable to the particle 
tracks the emission from all particles is coherent and the 
excess charge distribution in the shower generates a complex 
radiation pattern. It is most convenient to work directly 
with the Fourier transform of the radiated electric field, 
$\vec{E}$, which can be directly obtained from Maxwell's 
equations in a dielectric medium \cite{Allan,Jelley,Zas}. 
In the Fraunhofer limit (observation distance $R$ much 
greater than the tracklength) the contribution of 
an infinitesimal particle track $\vec{\delta l}=\vec{v} \delta t$ 
is given by:
\begin{equation}
R \vec E(\omega,{\vec {\rm x}})=
{e \mu_{\rm r}~i \omega \over 2 \pi \epsilon_0 {\rm c}^2}~
{\vec  \delta l}_{\perp} ~ 
{\rm e}^{i(\omega -{\vec k} {\vec v}_1){\rm t_1} } 
~{\rm e}^{ikR},
\label{deltat} 
\end{equation}
where $\vec k$, $\vec k \parallel \vec R$, is the wave vector in the direction
of observation ($\vec R$) and $\vec l_{\perp}$ is the tracklength 
projected onto a plane perpendicular to the 
observing direction. 

This simple expression displays in a transparent form three 
most important characteristics of such signals: The 
proportionality between the electric field amplitude and the 
tracklength, the fact that in the Cherenkov direction 
($\omega - \vec k \cdot \vec v=0$) there is no phase factor 
associated to the position along the track direction 
and the fact that radiation is polarized in the direction of 
$\vec l_{\perp}$, that is in the apparent direction of the track 
as seen from an observer located at $\vec {\rm x}$. 

\begin{figure}[hbt]
\centering
\vspace*{-2.5cm}
\mbox{\epsfig{figure=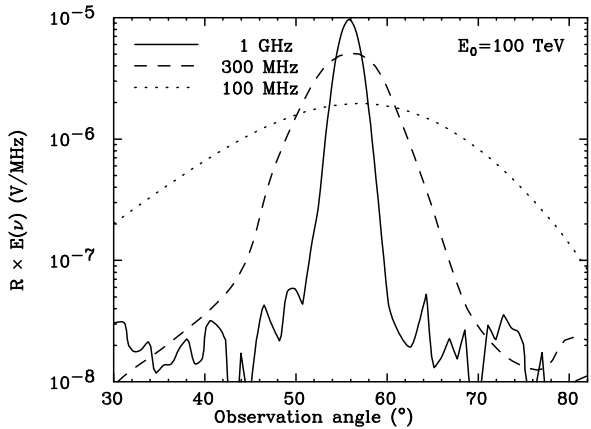,width=9.0cm}}
\vspace*{-3cm}  
\caption{Electric field angular distribution.}
\label{fig:prl1etheta}
\end{figure}

\begin{figure}[hbt]
\centering
\vspace*{-2.5cm}
\mbox{\epsfig{figure=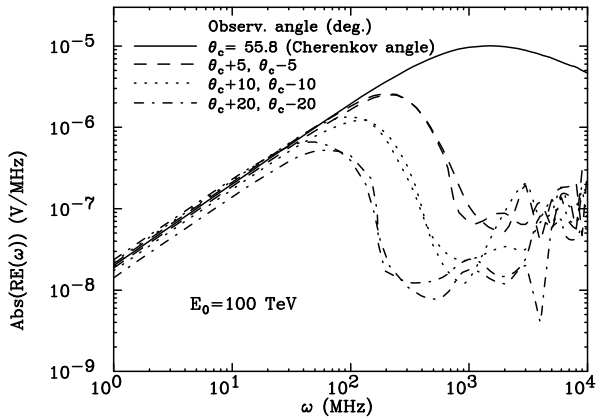,width=9.0cm}}
\vspace*{-3cm}  
\caption{Electric field frequency spectrum.}
\label{fig:prl2enu}
\end{figure}

Recent numerical simulations of radio pulses from 
both electromagnetic 
and hadronic showers \cite{Zas91,Alz97,Alz98}
are based on this expression. 
For energies below 10~PeV full simulations are possible 
\cite{Zas91}. The characteristic angular distributions and 
frequency spectra are shown in 
Figs.~\ref{fig:prl1etheta},\ref{fig:prl2enu}. 
We can understand most of the pulse characteristics by 
studying the particle distributions in a shower as the 
excess charge follows the electron distribution closely. 

To a good approximation the pulse is the Fourier 
transform of the spatial distribution of the excess 
charge. For many purposes it is sufficient to study 
the Fourier transform of the one dimensional 
distribution (in shower axis $z$) as has been extensively 
checked \cite{ZAl98inprep}:  
\begin{equation}
R \vert \vec E(\omega,{\vec {\rm x}}) \vert \simeq 
{e \mu_{\rm r}~i \omega \over 2 \pi \epsilon_0 
{\rm c}^2}~{\rm e}^{ikR}~sin\theta~ 
\int dz~Q(z) {\rm e}^{ipz}
\label{approx1d} 
\end{equation}
We here introduce the parameter 
$p(\theta,\omega)= \omega /c (1-n~\cos \theta)$ to  
transparently relate the radio emission spectrum 
to the Fourier transform of the (excess) charge distribution 
$Q(z)$. This approximation together with hybrid techniques 
combining simulation and parameterization of shower development 
have allowed the characterization of pulses from showers of 
energy up to 100~EeV \cite{Alz97,Alz98}. 

The angular distribution of the pulse has a main ''diffraction'' 
peak corresponding to $p=0$, the Cherenkov direction, see 
Fig.~\ref{fig:prl1etheta}. 
For $|p|~l_{sh} \ll 1$, where $l_{sh}$ is a length scale parameter 
for the shower \cite{Alz97}, the electric field 
spectrum accurately scales with electron tracklength, see 
Fig.~\ref{fig:prl2enu}.
In electromagnetic showers the tracklength is proportional to 
the energy\cite{Alz97} and for hadronic showers it scales 
with a slowly varying fraction of the energy 
($80-92~\%$ for shower energies 
between 100~TeV and 100~EeV) \cite{Alz98}. 

The scaling with electromagnetic energy is broken by 
interference from different parts of the shower when 
$|p|~l_{sh} \sim 1$. As a result the frequency spectrum 
stops rising linearly with frequency and has a maximum 
$\omega_{M}(\theta)$ which depends strongly on $\theta$ 
as seen in Fig.~\ref{fig:prl2enu}. Expanding the 
condition for $p(\theta)$ about $\theta_C$ it simply 
becomes 
$ n \sin \theta_C l_{sh} \Delta \theta \omega_M /c \sim 1$
which clearly displays how $\Delta \theta$ is inversely 
proportional to $\omega_M$ as shown in the figure.
This allows independent establishment of the angle between 
the observation and Cherenkov directions which is not
sufficient to establish the shower direction but it can be 
combined with other measurements to provide useful information. 
This relation however breaks down when approaching the 
Cherenkov direction because the lateral distribution plays 
the destructive role although there is no 
interference from different shower depths (Eq.~\ref{deltat}).

The ''central peak'' at 1~GHz concentrates 
most of the power. 
For given frequency the angular spread of the pulse is also 
inversely proportional to $l_{sh}$. This effect hardly shows 
in showers below 10~PeV with a longitudinal scale that 
only depends logarithmically on 
energy\cite{Greisen}. The difference between the longitudinal development 
of the excess charge  
for electromagnetic and hadronic showers is not enough to 
show up in the radiopulse structure (both are governed 
by the radiation length of the material). 
Nevertheless the angular width of the pulse reduces significantly  
for the characteristically elongated electromagnetic showers 
above 100~PeV because of the LPM effect \cite{LPMStanev}. 
This narrowing of the angular distribution allows the identification 
of elongated showers.   

The LPM effect manifests as a dramatic reduction of the pair 
production and bremsstrahlung cross sections at large energies 
due to large scale correlations in the atomic electric fields 
\cite{LPM,Klein}. It only affects the  development of  
showers initiated by photons, electrons or positrons above 
a given energy, about $20$~PeV in ice 
\cite{LPMStanev}. Showers initiated by EeV hadrons 
have high multiplicities (50-100) in their first interaction, 
and the pions produced typically have energies $1-2\%$ 
that of the initial hadron. Moreover $\pi^0$'s above 
6.7~PeV are more likely to interact in ice than to decay 
and only about $2\%$ of the hadron showers above 10 EeV 
have one photon with more than $10\%$ of the hadron energy.
Furthermore in a 100 EeV neutrino interaction for example 
the fraction of energy transferred to the hadron debris 
($25\%$ in average) fragments into about 17 hadrons 
(mostly pions) which have about $5\%$ of the transferred energy 
(except for the leading baryon which would carry a fraction $1-K$ 
where $K$ is the inelasticity). As a result the photons that 
are responsible for the electromagnetic subshowers  
(from $\pi^0$ and other short lived particles decay) 
have energies which are far removed from that of the 
initial neutrino. Very few hadronic showers 
induced by neutrino interactions of 100~EeV would display 
an LPM tail. For the photon to exceed $100$~PeV 
with a probability greater than $2\%$, the initial 
neutrino energy should be above 80~EeV. 

\begin{figure}[hbt]
\centering
\vspace*{-2.5cm}
\mbox{\epsfig{figure=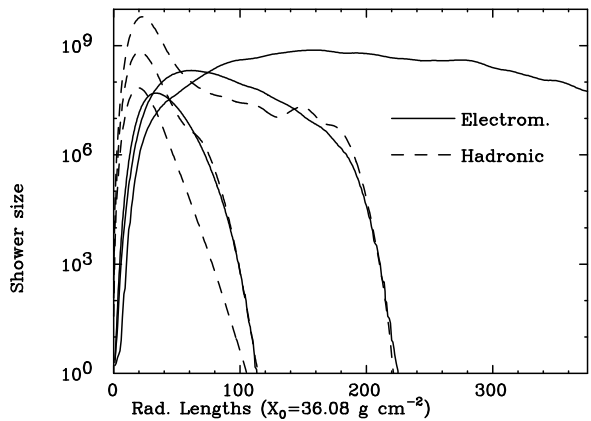,width=9.0cm}}
\vspace*{-3cm}  
\caption{Longitudinal development of electromagnetic and hadronic showers.}
\label{fig:prl3em-had}
\end{figure}

\begin{figure}[hbt]
\centering
\vspace*{1.cm}
\mbox{\epsfig{figure=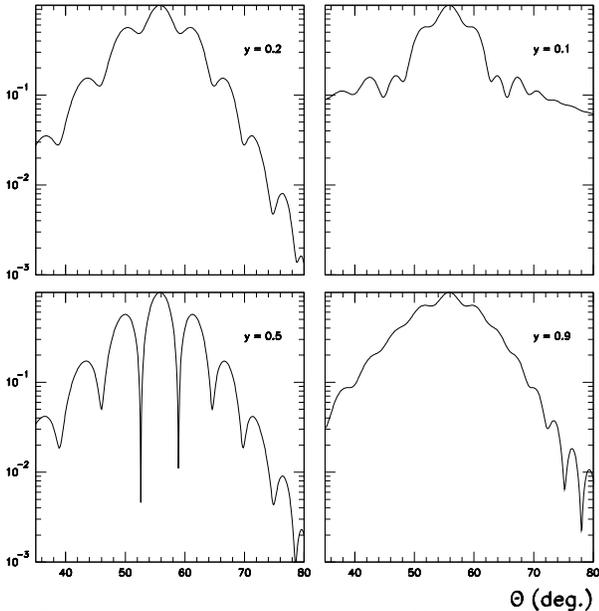,width=9.0cm}}
\vspace{-2.5cm}
\caption{Interference pattern for $\nu = 100$ MHz as obtained in charged current neutrino interactions with different energy transfers to the hadron ($y$) 
as indicated.}
\label{fig:prl4y-pulse}
\end{figure}

The elongation has a dramatic effect on the angular 
distribution of the radio pulse. For electromagnetic showers 
the central peak width narrows as $E^{-1/3}$ above 
20~PeV\cite{Alz97}. A 10~EeV electron produces a pulse which is 
about 10 times narrower than that of a hadronic shower 
of the same energy what makes 
differentiation between pulses from electromagnetic and 
hadronic cascades possible in principle, allowing the 
characterization of electron neutrinos (see Fig.~\ref{fig:prl3em-had}). 
For showers initiated by hadrons above 10~EeV the pulse shows a 
characteristic angular distribution of interference 
of two periodicities corresponding to the two length scales, 
one associated to the hadronic shower and the second, longer 
but of much less intensity associated to the electromagnetic 
LPM tail \cite{Alz98}. The radio pulse for an electron neutrino 
interaction has an interference pattern of similar nature. 
As the average fraction of energy transfer to the hadron debris is expected to 
be about $<y>=0.25$ \cite{Gonza98} this interference effect is 
typically enhanced as shown in Fig.~ \ref{fig:prl4y-pulse}. The 
angular distribution of the pulse retains enough information to 
allow independent extraction of the total electromagnetic energy 
in both showers, that is to determine the individual energy transfer of the 
reaction. 


Typical energy thresholds for detecting electromagnetic showers 
with single antennas have been made in\cite{Zas,Alz97,Alz98,provorov} 
and are typically tens of PeV for showers produced 
at 1~km from the antenna assuming nominal frequencies of 
$\nu_0 = 1$~GHz and bandwidths of $0.1~\nu_0$. 
This corresponds of course to the case 
that the antenna lies just in the illuminated region of 
the central peak.
The volume of the illuminated region decreases linearly as $\nu _0$ rises and 
is also significantly different for electromagnetic and hadronic showers of 
energy above 100 PeV. 

Clearly a signal from a single antenna would be of little use for 
neutrino detection unless information about the shower 
direction and/or the shower energy could be obtained from it.  
If this was not the case it would be impossible to distinguish  
them from nearby pulses produced by low energy showers such as those 
induced by deeply penetrating muons. 
Information on neutrino interactions 
can only be obtained by placing antennas in an array 
covering a large region whether on the 
ice surface or under it \cite{ralston}.
The arrival times for pulses, the polarization \cite{Allan}, 
the relative amplitudes of the signals, and the
frequencies at different positions of the array 
elements are in principle experimentally accessible 
and would give relevant and redundant information.
The technique is similar to ''conventional'' neutrino detector proposals 
but can be highly enriched with the angular diffraction patterns, 
the frequency spectra and the polarization. 

For intermediate energies one looks for events coming from "below" 
where the Earth provides a shield for all other types of particles. 
For extremely high energies 
($>$ PeV) however the Earth becomes opaque and neutrino 
events have to be searched in the horizontal direction or possibly 
from "above". Although some high energy showers can be expected from 
other processes such as atmospheric muon bremsstrahlung at PeV energies 
the background of these events is sufficiently suppressed. 
There is redundant information that allows a variety of cross checks. 
For instance timing can be used to establish the shower position in a manner 
very similar to arrays of particle detectors detecting air showers, 
but the spatial distribution of the signal in the detector can 
be also used for the same purpose, even signal polarization 
provides an interesting cross check of the shower orientation, 
what will also be particularly useful to filter spurious noise signals out.   

The antenna/array parameters are crucial for performance. Most importantly 
operating frequency 
$\nu_0$, bandwidth $\Delta \nu$ and array spacing. These parameters 
are deeply interconnected at detection level and therefore require 
complex optimization once the noise levels are well understood. 
The spacing of the antenna array will determine the minimum distance at 
which the geometry of the illuminated region corresponding to the 
Cherenkov central peak can be reconstructed, what would indicate the 
position of the shower\cite{markov}. The nominal 
frequency will determine both the width of the diffraction peaks and the 
transmitting properties of the medium so that array spacing 
should be adjusted to the choice of antenna. 
Lastly the larger the bandwidth the better the signal to noise 
ratio because the noise behaves as $\sqrt{\Delta \nu}$. 

Neglecting attenuation the ratio of shower energy to distance 
has to be above a given value for a shower to be detected. 
As $\nu_0$ approaches the naively optimal value 
for coherence of 1~GHz the attenuation distance drops below 
1~km and what is more problematic 
temperature effects become important\cite{Allan} 
possibly forcing detections to be 
within scales less than $1$~km for such antennas. 
Using lower frequencies the signal to noise ratio drops and higher 
energy thresholds are needed to compensate the loss of signal. 
This may be nevertheless advisable if one is ready to concentrate on 
neutrinos of the highest energies allowing detection at distances above 
1~km. 

We have discussed the implications of radio pulse calculations 
for high energy shower detection stressing how different 
features of the signal can be used for shower 
characterization. We have shown how the LPM effect allows 
the separation of charged current electron neutrino interactions 
from the rest, and in principle how the technique can be used to 
extract the energy fraction transmitted to the electron. 
We have avoided the discussion of unresolved 
experimental issues \cite{Allan,zeleznihk}, i.e. noise, which 
are likely to determine the final sensitivity of the technique, 
that is the precise energy value above which showers become 
detectable over sufficiently long distances. This sensitivity 
will be also completely dependent on the experimental setup 
which will have to be optimized accordingly. These crucial 
issues have to be addressed with in situ experiments and there 
are efforts in this direction \cite{RICE}, but are unlikely to 
change the general conclusions obtained here.  
 
\vskip 0.5 cm
{\bf Acknowledgements:} We thank F.~Halzen for suggestions after reading the 
manuscript and G.~Parente, T. Stanev and I.M. Zheleznykh for helpful 
discussions. This work was supported in part by CICYT (AEN96-1773) and by 
Xunta de Galicia (XUGA-20604A96). J.A. thanks the Xunta de Galicia for 
financial support.


\begin{thebibliography}{999}
\bibitem{physreps} T.K. Gaisser, F. Halzen, and T. Stanev, Phys. Rep. 
{\bf 258} (1995) 173 and references therein.
%
\bibitem{icecube} F. Halzen, {\it The case for a kilometer-scale neutrino detector}, in Nuclear  
and Particle Astrophysics and Cosmology, Proceedings of Snowmass\,94,  
R.\,Kolb and R.\,Peccei, eds.; {\it The Case for a Kilometer-Scale Neutrino  
Detector: 1996}, in Proc.\ of the Sixth International Symposium on Neutrino  
Telescopes, ed.\ by  M.\,Baldo-Ceolin, Venice (1996). 
%
\bibitem{oscillations} Y. Fukuda {\it et al.}, Phys. Rev. Lett. {\bf 81} 
(1998) 1562.
%
\bibitem{longoscil} H. Minakata, A. Yu. Smirnov, Phys. Rev. {\bf D54} (1996)
3698.
%
\bibitem{Cronin} K.S. Capelle, J.W. Cronin, G. Parente, and E. Zas,
Astro. Phys. {\bf 8} (1998) 321. 
%
\bibitem{askaryan} G.A. Askar'yan, Soviet Physics JETP {\bf 14} (1962)2 441;
{\bf 48} (1965) 988.
%
\bibitem{markov} M.A. Markov and I.M. Zheleznykh, Nucl. Instr. and Methods in 
Phys. Research {\bf A248} (1986) 242.
%
\bibitem{price} P.B. Price, Astro. Phys. {\bf 5} (1996) 43.
%
\bibitem{fran} I. Frank, I. Tamm, {\sl Dokl.\ Akad.\ Nauk\ SSSR} {\bf 14}
109 (1937).
%
\bibitem{ralston} G.M. Frichter et al.,  Phys. Rev. 
{\bf D53},3 (1996), 1684.
%
\bibitem{Allan} H.R. Allan, {\sl Progress in Elementary Particles and Cosmic 
Ray Physics} (North Holland, 1971), Vol. 10, p. 171.
%
\bibitem{Jelley} J.V. Jelley, Astro. Phys. {\bf 5} (1996) 255 and 
refs. therein.
%
\bibitem{Zas} E. Zas, F. Halzen, T. Stanev, Phys. Rev. {\bf D45} (1992) 362.
%
\bibitem{Zas91} F.Halzen, E.Zas, T.Stanev, Phys. Lett. {\bf B257} 1991 432.
%
\bibitem{provorov} A.L. Provorov, I.M. Zheleznykh, Astro. Phys. {\bf 4}
(1995) 55.
%
\bibitem{Alz97} J. Alvarez-Mu\~niz, E. Zas, Phys. Lett. {\bf B411} 218 
1997.
%
\bibitem{Alz98} J. Alvarez-Mu\~niz, E. Zas, Phys. Lett. {\bf B434}
396 (1998).
%
\bibitem{ZAl98inprep} J. Alvarez-Mu\~niz and E. Zas, work in preparation.
%
\bibitem{Greisen} K. Greisen, Prog. of Cosmic Ray Phys., ed.
J.G. Wilson, Vol.~III, (North Holland, 1956) p.1.
%
\bibitem{LPMStanev}  T. Stanev et al. 
Phys. Rev. {\bf D25} (1982) 1291.
%
\bibitem{LPM} L. Landau and I. Pomeranchuk, {\sl Dokl.\ Akad.\ Nauk\ SSSR}
{\bf 92} (1953) 535; {\bf 92} (1935) 735; A.B. Migdal, Phys. Rev.
{\bf 103} (1956) 1811;  Sov. Phys. JETP {\bf 5} (1957) 527.
%
\bibitem{Klein} S.R. Klein, hep-ph/9802442, to be publ. in Rev. Mod. Phys; 
also astro-ph/9712198.
%
\bibitem{Gonza98} J. Castro, G. Parente, and E. Zas, in preparation.
%
\bibitem{zeleznihk} I.M. Zheleznykh, {\sl Proc.\ XXIth ICRC} (Adelaide, 1989), 
Vol.~6, p.~528.
%
\bibitem{RICE} C. Allen {\it et al.} {\sl Proc. High Energy Physics Conf.} 
1998.
%
\end{thebibliography}
\end{document}